\documentclass[a4paper,nobibnotes,nofootinbib,preprintnumbers]{revtex4}
\usepackage{amssymb}
\usepackage{amsmath}
\usepackage{epsfig}
\newcommand{\be}{\begin{equation}}
\newcommand{\ee}{\end{equation}}
\newcommand{\bea}{\begin{eqnarray}}
\newcommand{\eea}{\end{eqnarray}}
\newcommand{\nn}{\nonumber}

\newcommand{\D}{\displaystyle}
\newcommand{\g}{\gamma}
\newcommand{\f}{\frac}

\newcommand{\intc}[1]{{\int\frac{d#1}{2i\pi}}}
\newcommand\lr[1]{{\left({#1}\right)}}

\begin{document}
\title{Next-to-leading BFKL phenomenology of forward-jet cross sections at HERA}
\author{O. Kepka}\email{oldrich.kepka@cea.fr}
\affiliation{DAPNIA/Service de physique des particules, CEA/Saclay, 91191 
Gif-sur-Yvette cedex, France}
\author{C. Marquet}\email{marquet@quark.phy.bnl.gov}
\affiliation{RIKEN BNL Research Center, Brookhaven National Laboratory, Upton, NY 
11973, USA}
\author{R. Peschanski}\email{pesch@spht.saclay.cea.fr}
\affiliation{Service de physique th{\'e}orique, CEA/Saclay, 91191 
Gif-sur-Yvette cedex, France
\\URA 2306, unit{\'e} de recherche associ{\'e}e au CNRS}
\author{C. Royon}\email{royon@hep.saclay.cea.fr}
\affiliation{DAPNIA/Service de physique des particules, CEA/Saclay, 91191 
Gif-sur-Yvette cedex, France}

\preprint{SPhT-T06/187}
\preprint{RBRC-632}

\begin{abstract}

We show that the forward-jet measurements performed at HERA allow for a detailed
study of corrections due to next-to-leading logarithms (NLL) in the 
Balitsky-Fadin-Kuraev-Lipatov (BFKL) approach. While the description of the
$d\sigma/dx$ data shows small sensitivity to NLL-BFKL corrections, these can be 
tested by the triple differential cross section $d\sigma/dxdk_T^2dQ^2$ recently 
measured. These data can be successfully described using a renormalization-group 
improved NLL kernel while the standard next-to-leading-order QCD or 
leading-logarithm BFKL approaches fail to describe the same data in the whole 
kinematic range. We present a detailed analysis of the NLL scheme and 
renormalization-scale dependences and also discuss the photon impact factors.

\end{abstract}
\maketitle
\section{Introduction}

Forward-jet production in lepton-proton deep inelastic scattering is a process in 
which a jet is detected at forward rapidities in the direction of the proton. 
This process is characterized by two hard scales: $Q^2,$ the virtuality of the 
intermediate photon that undergoes the hadronic interaction and $k_T^2,$ the 
squared transverse momentum of the forward jet. When the total energy of the 
photon-proton collision $W$ is sufficiently large, corresponding to a small value 
of the Bjorken variable $x\!\simeq\!Q^2/W^2,$ forward-jet production is relevant 
\cite{mueller} for testing the Balitsky-Fadin-Kuraev-Lipatov (BFKL) approach 
\cite{bfkl}.

In fixed-order perturbative QCD calculations, the hard cross section is computed 
at fixed order with respect to $\alpha_s,$ and large logarithms coming from the 
strong ordering between the proton scale and the forward-jet scale are resummed 
using the Dokshitzer-Gribov-Lipatov-Altarelli-Parisi (DGLAP) evolution equation 
\cite{dglap}. However in the small$-x$ regime, other large logarithms arise in 
the hard cross section itself, due to the strong ordering between the energy $W$ 
and the hard scales. These can be resummed using the BFKL equation, at leading 
(LL) and next-leading (NLL) logarithmic accuracy \cite{bfkl,nllbfkl}. 

It has been shown that the H1 and ZEUS $d\sigma/dx$ forward-jet data
\cite{h1zeus99,h1new,zeusnew} are well described by LL-BFKL predictions 
\cite{fjetsll,fjetsll2}, while fixed-order perturbative QCD predictions at
next-to-leading order (NLOQCD) fail to describe the data, underestimating the 
cross section by a factor of about 2 at small values of $x.$ However, these tests 
on the relevance of BFKL dynamics have not been considered fully conclusive. On 
the theoretical side, it has been found that NLL-BFKL corrections \cite{nllbfkl} 
could be large enough to invalidate the tests. On a phenomenological side, other 
models such as DGLAP evolution with a ``resolved'' photon \cite{respho} could 
increase the NLOQCD predictions and come to reasonable agreement with the data. 

The recent experimental forward-jet measurements \cite{zeusnew,h1new} performed at 
HERA motivate a new phenomenological analysis of BFKL effects in forward-jet 
cross sections.
In particular the triple differential cross section $d\sigma/dxdk_T^2dQ^2,$
allows for a detailed study of the QCD dynamics of forward jets. Contrary to the 
$d\sigma/dx$ data, which were obtained with kinematical cuts such that
$r\!=\!k_T^2/Q^2\!\sim\!1,$ the triple differential cross section is measured 
with different sets of cuts such that the data are also sensitive to the regime 
$r\!\gg\!1,$ where the two hard scales of the problem are somewhat ordered. While 
LL-BFKL predictions describe well the data obtained with $r\!\sim\!1,$ it was 
noticed \cite{fjetsll2} that they fail to describe the $r\!\gg\!1$ regime, 
indicating the need for NLL-BFKL corrections.

It was known that NLL-BFKL corrections could be large due to the appearance of 
spurious singularities in contradiction with renormalization-group requirements. 
However it has been realized \cite{salam,CCS} that a renormalization-group improved 
NLL-BFKL regularization can solve the singularity problem and lead to reasonable 
NLL-BFKL kernels (see also \cite{singnll} for different approaches). This motivates 
the present 
phenomenological study of NLL-BFKL effects in forward-jet production. Even though 
the determination of the next-leading impact factors is still in progress 
\cite{nllif}, our analysis allows us to study the NLL-BFKL framework, and 
the remaining ambiguity corresponding to the dependence on the specific 
regularization scheme. Our goal is to confront the new experimental data, in 
particular the triple-differential cross section, to NLL-BFKL predictions in 
different schemes.

In Ref.\cite{nllf2}, such a phenomenological investigation has been devoted to 
the proton structure function data, taking into account NLL-BFKL effects through 
an ``effective kernel'' (introduced in \cite{CCS}) using different schemes. A 
saddle-point approximation for hard enough scales was used to evaluate the BFKL 
Mellin integration which allowed one to obtain a phenomenological description of 
NLL-BFKL effects. In the present study devoted to forward-jet production, we take 
into account the proper symmetric two-hard-scale feature of the forward-jet problem 
when introducing the effective kernel, and we implement the NLL-BFKL effects with 
an exact Mellin integration, rather than a saddle-point approximation.

Some preliminary results, mostly based on the saddle-point approach, were presented 
in
\cite{us}. They showed the potential of forward-jet data on $d\sigma/dx$ and 
specially
$d\sigma/dxdk_T^2dQ^2$ to discuss NLL effects in the BFKL approach. In this paper, 
we systematically use an exact Mellin integration, and we present a detailed 
analysis of the NLL scheme and scale dependences and also discuss the sensitivity 
of our NLL-BFKL descriptions with respect to the photon impact factors. We also 
study the NLOQCD predictions, testing their relevance by comparing the use of 
different parton densities and different renormalization and factorization scales.

The plan of the paper is the following. In section II, we present the 
phenomenological NLL-BFKL formulation of the forward-jet cross section for the two 
schemes called S3 and S4, while briefly highlighting the principles of its 
derivation. In section III, we compare the predictions of the two NLL-BFKL schemes 
with the data, and also with LL-BFKL and NLOQCD predictions. We discuss the dependence of 
our results on the choice of the hard scale with which $\alpha_s$ is 
running in Section IV, and on the assumption made for the photon 
impact factors in Section V. Section VI presents the scale and parton-density 
dependences of the NLOQCD predictions. Section VII is devoted to conclusions and an
outlook.

\section{Forward-jet production in the BFKL framework}

Forward-jet production in a lepton-proton collision is represented in Fig.1 
with the different kinematic variables. We denote by $\sqrt{s}$ the total energy of 
the lepton-proton collision and by $Q^2$ the virtuality of the intermediate photon 
that undergoes the hadronic interaction. We shall use the usual kinematic 
variables of deep inelastic scattering: $x\!=\!Q^2/(Q^2\!+\!W^2)$ and 
$y\!=\!Q^2/(xs)$ where $W$ is the center-of-mass energy of the photon-proton 
collision. In addition, $k_T\!\gg\!\Lambda_{QCD}$ is the transverse momentum of the jet 
and $x_J$ its longitudinal momentum fraction with respect to the proton. The QCD 
cross section for forward-jet production reads
\be
\f{d^{(4)}\sigma}{dxdQ^2dx_Jdk_T^2}=\f{\alpha_{em}}{\pi xQ^2}
\left\{\lr{1-y+\f{y^2}2}\f{d\sigma^{\g*p\!\rightarrow\!JX}_T}{dx_Jdk_T^2}+
(1-y)\f{d\sigma^{\g*p\!\rightarrow\!JX}_L}{dx_Jdk_T^2}\right\}\ 
,\label{fj}\ee
where $d\sigma^{\g*p\!\rightarrow\!JX}_{T,L}/dx_Jdk_T^2$ is the cross section 
for forward-jet production in the collision of the transversely (T) or 
longitudinally (L) polarized virtual photon with the target proton.

In the following, we consider the high-energy regime $x\!\ll\!1$ in which the 
rapidity interval $Y\!=\!\log(x_J\!/\!x)$ is assumed to be very large. The NLL-BFKL
forward-jet cross section of our analysis is given by:
\be
\f{d\sigma^{\g*p\!\rightarrow\!JX}_{T,L}}{dx_Jdk_T^2}=
\f{\alpha_s(k_T^2)\alpha_s(Q^2)}{k_T^2Q^2}\ f_{eff}(x_J,k_T^2)
\intc{\g}\lr{\f{Q^2}{k_T^2}}^\g \phi^\g_{T,L}(\g)\ 
e^{\bar\alpha(k_T Q)\chi_{eff}[\g,\bar\alpha(k_T Q)]Y}
\label{nll}
\ee
with the complex integral running along the 
imaginary axis from $1/2\!-\!i\infty$ to $1/2\!+\!i\infty.$ The running coupling is
given by
\be
\bar\alpha(k^2)=\alpha_s(k^2)N_c/\pi=
\left[b\log\lr{k^2/\Lambda_{QCD}^2}\right]^{-1}\ ,
\hspace{0.5cm}\mbox{with}\hspace{0.5cm}b=\f{11N_c-2N_f}{12N_c}\ .\label{runc}\ee
In formula \eqref{nll}, the NLL-BFKL effects are phenomenologically taken
into account by the effective kernel $\chi_{eff}(\g,\bar\alpha).$ Let us now give
further details on this approximation.

\begin{figure}[t]
\begin{center}
\epsfig{file=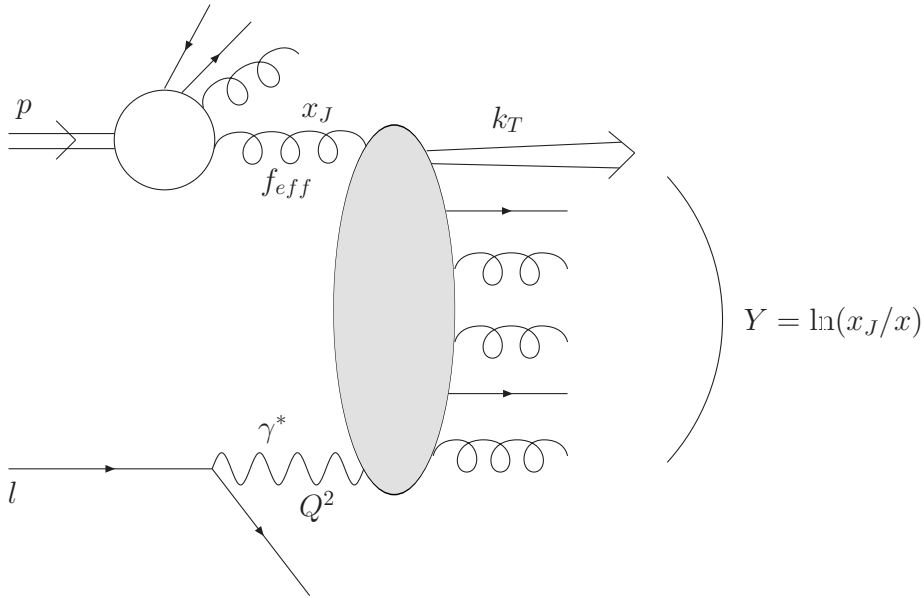,width=12cm}
\caption{Production of a forward jet in a lepton-proton collision. The kinematic 
variables of the problem are displayed. $Q^2$ is the virtuality of the photon 
that undergoes the hadronic interaction. $k_T$ is the transverse momentum of the 
forward jet and $x_J$ is its longitudinal momentum fraction with respect to the 
incident proton. $Y$ is the (large) rapidity interval between the two hard probes.}
\end{center}
\end{figure} 

The scheme-dependent NLL-BFKL kernels provided by the regularization procedure
$\chi_{NLL}\lr{\g,\omega}$ depend on $\g,$ the Mellin variable conjugate to 
$Q^2/k_T^2$ and $\omega,$ the Mellin variable conjugate to $W^2/Qk_T.$ In this work
we shall consider the S3 and S4 schemes \cite{salam}, recalled in Appendix A, in which
$\chi_{NLL}$ is supplemented by an explicit $\bar\alpha$ dependence. One writes the following {\it consistency condition} \cite{salam,ccond}
\be
\omega=\bar\alpha\ \chi_{NLL}\left(\g,\omega\right)
\label{consistency}
\ee
which represents the diagonalized form of the NLL-BFKL evolution equation and allows one to formulate the cross section \eqref{nll} in terms of $\chi_{eff}(\g,\bar\alpha).$
The approximation amounts to introduce the effective kernel to satisfy the 
consistency condition. Indeed, the effective kernel $\chi_{eff}(\g,\bar\alpha)$ is
defined from the NLL kernel $\chi_{NLL}\lr{\g,\omega}$ by solving the implicit equation
\be
\chi_{eff}(\g,\bar\alpha)=\chi_{NLL}\left[\g,\bar\alpha\ 
\chi_{eff}(\g,\bar\alpha)\right]\ ,
\label{eff}
\ee
as a solution of the consistency condition \eqref{consistency}.

To highlight how the effective kernel enters in the formulation of the forward-jet cross section, let us consider the following inverse Mellin transformation over $\omega,$ the variable conjugate to $e^Y$ ($\sim$ the energy squared), where $I(\g,\omega)$ represents next-to-leading order corrections to the LO impact factors $\phi^\g:$
\be
\intc{\omega}\f{I(\g,\omega)\ e^{\omega 
Y}}{\omega-\bar\alpha\chi_{NLL}(\g,\omega)}=
\f{I[\g,\bar\alpha\chi_{eff}\lr{\g,\bar\alpha}]}
{1-\dot\chi_{NLL}[\g,\bar\alpha\chi_{eff}\lr{\g,\bar\alpha}]}\ 
e^{\bar\alpha\chi_{eff}\lr{\g,\bar\alpha}Y}\ 
,\hspace{0.5cm}\mbox{with}\hspace{0.5cm}
\dot\chi_{NLL}=\f{d\chi_{NLL}}{d\omega}\ .
\label{omegaint}
\ee
The factor in front of the exponential is an unknown correction, due both to the 
yet unknown next-to-leading order corrections to the LO impact factors and to the 
approximations made in satisfying the consistency equation through the effective 
kernel method. For simplicity, in the $\gamma$ integration of \eqref{nll}, we 
choose to factor this term out and treat it as a constant normalization parameter.

Some other comments are in order.
\begin{itemize}
\item In formula \eqref{nll}, the renormalization scale is $k^2\!=\!k_T Q,$ in 
agreement with the energy scale \cite{renscal}. In practice, one solves \eqref{eff} with 
$\bar\alpha=\bar\alpha(k_T Q).$ Therefore, to each renormalization scale corresponds an effective kernel \cite{nllf2}. In Section IV, we shall test the 
sensitivity of our results when using $k^2\!=\!\lambda\ k_TQ$ and varying
$\lambda.$ Following formula \eqref{eff}, the effective kernel is modified 
accordingly for each scheme, and we also modify the energy scale
$k_T Q\!\rightarrow\!\lambda\ k_TQ.$
\item As we pointed out already, in formula \eqref{nll} we use the leading-order 
(Mellin-transformed) impact factors
\be
\lr{\begin{array}{cc}
\phi^\g_{T}(\g)\\ \phi^\g_{L}(\g)
\end{array}}
=\pi\alpha_{em}N_c^2\sum_q
e_q^2\f{1}{2\g^2}
\f{\Gamma^3(1+\g)\Gamma^3(1-\g)}
{\Gamma(2-2\g)\Gamma(2+2\g)(3-2\g)}
\lr{\begin{array}{cc}(1+\g)(2-\g)\\2\g(1-\g)\end{array}}\ .
\label{phig}
\ee
for a transversely (T) and longitudinally (L) polarized virtual photon where 
$e_f$ is the charge of the quark with flavor $f.$ We consider massless quarks and 
sum over four flavors in \eqref{phig}. This is justified considering the rather 
high values of the photon virtuality ($Q^2\!>\!5\ \mbox{GeV}^2$) used for the 
measurement. We point out that our phenomenological approach can be adapted to 
full NLL accuracy, once the next-to-leading order impact factors are available
(the jet impact factors are known at next-to-leading order \cite{ifnlo}). For 
completeness, we shall discuss the sensitivity of our results to 
typical next-leading modifications of $\phi^\g_{T,L}(\g)$ in Section V.
\item In formula \eqref{nll}, $f_{eff}(x_J,k_T^2)$ is the effective parton 
distribution function and resums the leading logarithms 
$\log(k_T^2/\Lambda_{QCD}^2).$ It obeys the following expression
\be
f_{eff}(x_J,k_T^2)=g(x_J,k_T^2)+
\f{C_F}{N_c}\lr{q(x_J,k_T^2)+\bar{q}(x_J,k_T^2)}\ ,
\label{sf}\ee
where $g$ (resp. $q$, $\bar{q}$) is the gluon (resp. quark, antiquark) 
distribution function in the incident proton. Since the forward-jet measurement 
involves perturbative values of $k_T$ and moderate values of $x_J,$ formula 
\eqref{nll} features the collinear factorization of $f_{eff},$ with $k_T^2$ 
chosen as the factorization scale.
\item By comparison, the LL-BFKL formula is formally the same as \eqref{nll}, 
with the substitutions
\be
\chi_{eff}\rightarrow\chi_{LL}(\g)=2\psi(1)-\psi(1-\g)-\psi(\g)\ ,
\hspace{1cm}\bar\alpha(k^2)\rightarrow\bar\alpha=\mbox{const. parameter}\ ,
\label{nlltoll}\ee
where $\psi(\g)\!=\!d\log\Gamma(\g)/d\g$ is the logarithmic derivative of the 
Gamma function. One obtains
\be
\f{d\sigma^{\g*p\!\rightarrow\!JX}_{T,L}}{dx_Jdk_T^2}=
\f{\alpha_s(k_T^2)\alpha_s(Q^2)}{k_T^2Q^2}\ f_{eff}(x_J,k_T^2)
\intc{\g}\lr{\f{Q^2}{k_T^2}}^\g \phi^\g_{T,L}(\g)\ 
e^{\bar\alpha\chi_{LL}(\g)Y}
\label{ll}
\ee
\end{itemize}

\begin{figure}[t]
\begin{minipage}[t]{88mm}
\centerline{\epsfxsize=7.9cm\epsfbox{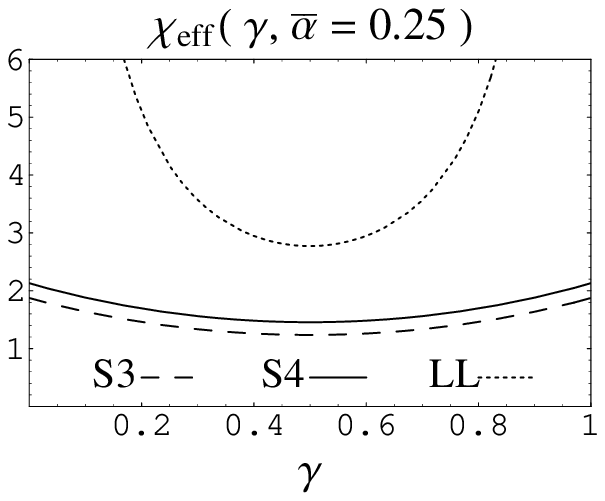}}
\end{minipage}
\hspace{\fill}
\begin{minipage}[t]{88mm}
\centerline{\epsfxsize=8.8cm\epsfbox{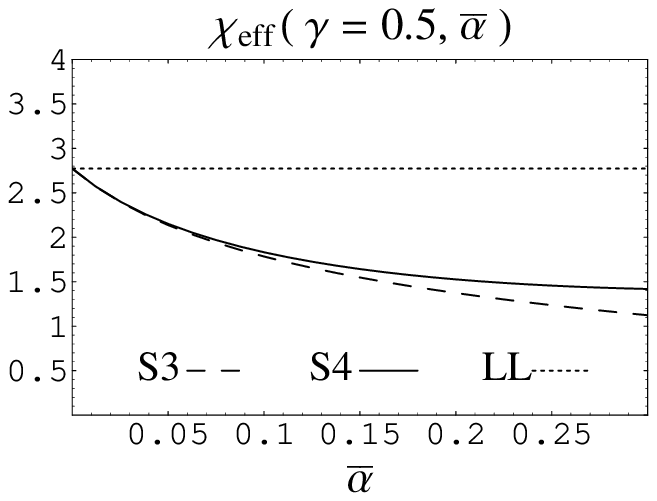}}
\end{minipage}
\caption{Comparison between the NLL-BFKL effective kernels $\chi_{eff}(\g,\bar\alpha)$ obtained by solving the implicit equation \eqref{eff} in the S3 and S4 schemes. The left plot shows 
$\chi_{eff}(\g,0.25)$ as a function of $\g$ while the right plot shows the minimum
$\chi_{eff}(0.5,\bar\alpha)$ as a function of $\bar\alpha.$ The dotted curves show the LL-BFKL kernel $\chi_{LL}(\g).$}
\end{figure}

Inserting formula \eqref{nll} (resp. \eqref{ll}) into \eqref{fj} 
gives the forward-jet cross section in the NLL-BFKL (resp. LL-BFKL) energy 
regime. In the LL-BFKL case, this is a 2-parameter formula: the overall 
normalization and $\bar\alpha.$ In the NLL-BFKL case, each set of scales 
($Q^2,k_T^2$) defines the running coupling constant and therefore we deal with only 
one free parameter, the overall normalization. The interesting property of our 
phenomenological approach is that formula
\eqref{nll} has formally the structure of the LL formula, but with only one free 
parameter and a NLL kernel. The delicate aspect of the problem comes from the 
scheme-dependent effective kernel $\chi_{eff}.$ A comparison between the LL-BFKL 
kernel and the S3 and S4 effective kernels is shown in Fig.2.
As is well known, the 
NLL modifications to the BFKL kernel are quite important and will play an important 
phenomenological role in our analysis.

\section{NLL description of the H1 data}

The NLL-BFKL formula for the fully differential forward-jet cross section is 
obtained from \eqref{fj} and \eqref{nll}. To compare the corresponding 
prediction with the data, one has to carry out a number of integrations over the 
kinematic variables. They have to be done while properly taking into account the 
kinematic cuts applied for the different measurements. The procedure is the same as 
the one described in Ref.\cite{fjetsll2}, Appendix A. First one chooses the 
variables that lead to the weakest possible dependence of the differential cross 
section (we noticed that the best choice is $1/Q^2,$ $1/k_T^2,$ $\log(1/x_J),$ and 
$\log(1/x)$) and then the integrations are computed numerically following the 
experimental cuts defined in \cite{h1new,zeusnew}.

To fix the normalization (the only free parameter) and check the quality of the 
data description using the BFKL formalism, we start by fitting the $d\sigma/dx$ H1 
data \cite{h1new}. The choice of this data set corresponds to the kinematical 
domain where the BFKL formalism is expected to hold ($x\!\ll\!1$ and 
$Q^2/k_T^2\!\sim\!1$). We then use the relative normalizations obtained between the 
different NLL BFKL calculations (S3 and S4) to make predictions for the triple 
differential cross section $d\sigma/dxdk_T^2dQ^2.$ For this first analysis, the 
coupling $\bar\alpha$ is running with the scale $k_T Q.$

\subsection{The cross section $d\sigma/dx$}

We considered two kinds of fits: the first one is performed using statistical and 
systematics errors and the second one with statistical errors only. The 
systematics errors are very much point-to-point correlated, and this is why it is 
important to perform the fits with statistical errors only. Ideally, one should use 
the statistical errors added in quadrature with the uncorrelated ones but this 
information is not available.

The fit results to the $d\sigma/dx$ H1 data are given in Table I. The $\chi^2$ 
values (per degree of freedom) of the fits performed using the full
(statistical and systematics) errors 
are quite good, always less than 1, for the two NLL-BFKL schemes we considered. 
This shows the possibility of describing the forward-jet cross section using the 
BFKL formalism at next-to-leading logarithmic accuracy. The fits using statistical
errors only (which assume implicitly that the systematics are maximally correlated which is close to reality) bring about more constraints and show interesting features. The S4 fit can
describe the data better ($\chi^2\!=\!10.0/5$ d.o.f.) whereas the S3 scheme
shows a higher value of $\chi^2$ ($\chi^2\!=\!29.5/5$). This indicates that the S4
scheme is favored.

The curves corresponding to the fit with statistical errors only are displayed in 
Fig.3 (upper plot), and they are compared with the LL-BFKL results taken from
\cite{fjetsll2}. We notice the tiny difference between the LL and NLL results
(the corresponding curves 
are barely distinguishable on the figure). This confirms that the data are 
consistent with the BFKL enhancement towards small values of $x.$ Contrary to the 
proton structure function $F_2,$ the forward-jet cross section $d\sigma/dx$ does 
not show large NLL-BFKL corrections, once the overall normalization fitted. This is 
due to the rather small value of the coupling $\bar\alpha\!\simeq\!0.16$ obtained 
in the LL-BFKL fit \cite{fjetsll2}, corresponding to an unphysically large 
effective scale.

We also present in Fig.3 the fixed order QCD calculation based on the DGLAP 
evolution of parton densities. The next-to-leading order (NLO) prediction of 
forward-jet cross sections is obtained using the NLOJET++ generator
\cite{nlojetDIS}. CTEQ6.1M \cite{cteq} parton densities were used, and the 
renormalization $\mu_r$ and the factorization scale $\mu_f$ were set equal to 
$\mu_r^2=\mu_f^2=Qk_t^{max},$ where $k_t^{max}$ corresponds to the maximal 
transverse momentum of forward jets in the event. The NLOQCD predictions do not 
describe the data at small values of $x,$ as they are lower by a factor of order 2. 
The sensitivity of these predictions to variations of the 
renormalization and factorization scales will be discussed in Section VI, as well 
as their dependence when using other parton densities.

The fit parameters obtained with statistical error only will be used in the 
following to make predictions for other observables, namely the triple 
differential cross section $d\sigma/dxdk_T^2dQ^2.$ The value of $\bar\alpha$ for
the LL-BFKL fit will be kept as well as the normalizations of the 
different NLL-BFKL calculations (S3 and S4).

\begin{table}[b]
\begin{center}
\begin{tabular}{|c||c|c|c|} \hline
 scheme & fit & $\chi^2/dof$ & $N$\\ 
\hline
\hline
S3 & stat. $+$ syst. & 1.15/5 & 1.591 $\pm$ 0.089 \\
S3 & stat. only & 29.5/5 & 1.640 $\pm$ 0.019 $\pm$ 0.204 \\
\hline
S4 & stat. $+$ syst. & 0.48/5 & 1.356 $\pm$ 0.076 \\
S4 & stat. only & 10.0/5 & 1.374 $\pm$ 0.016 $\pm$ 0.172 \\
\hline
\end{tabular}
\end{center}
\caption{Results of the NLL-BFKL fits to the H1 $d\sigma/dx$ data. The relative 
values between the different overall normalizations $N$ can be compared.}
\end{table}

\clearpage
\begin{figure}[t]
\begin{center}
\epsfig{file=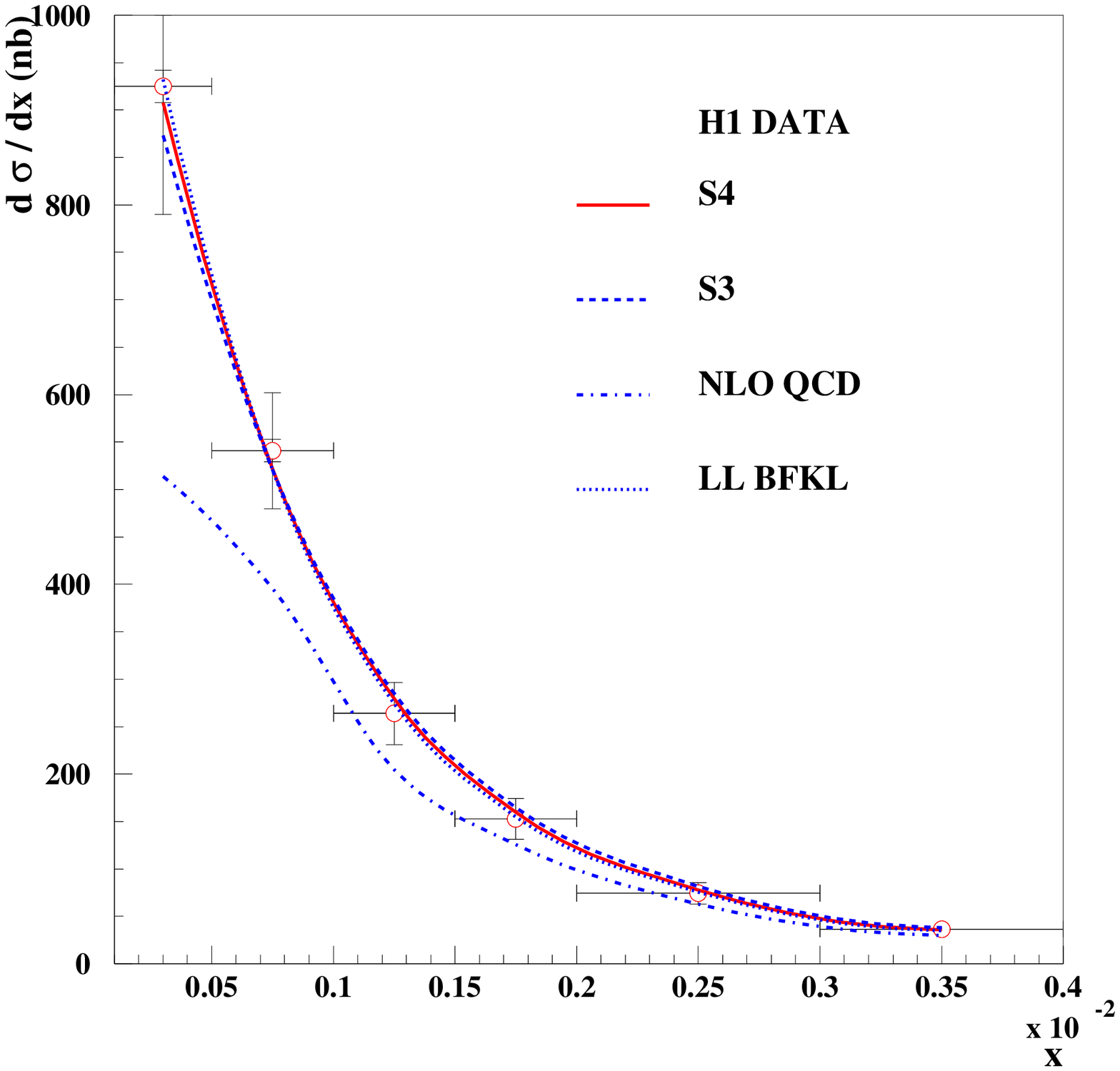,width=8.5cm}
\epsfig{file=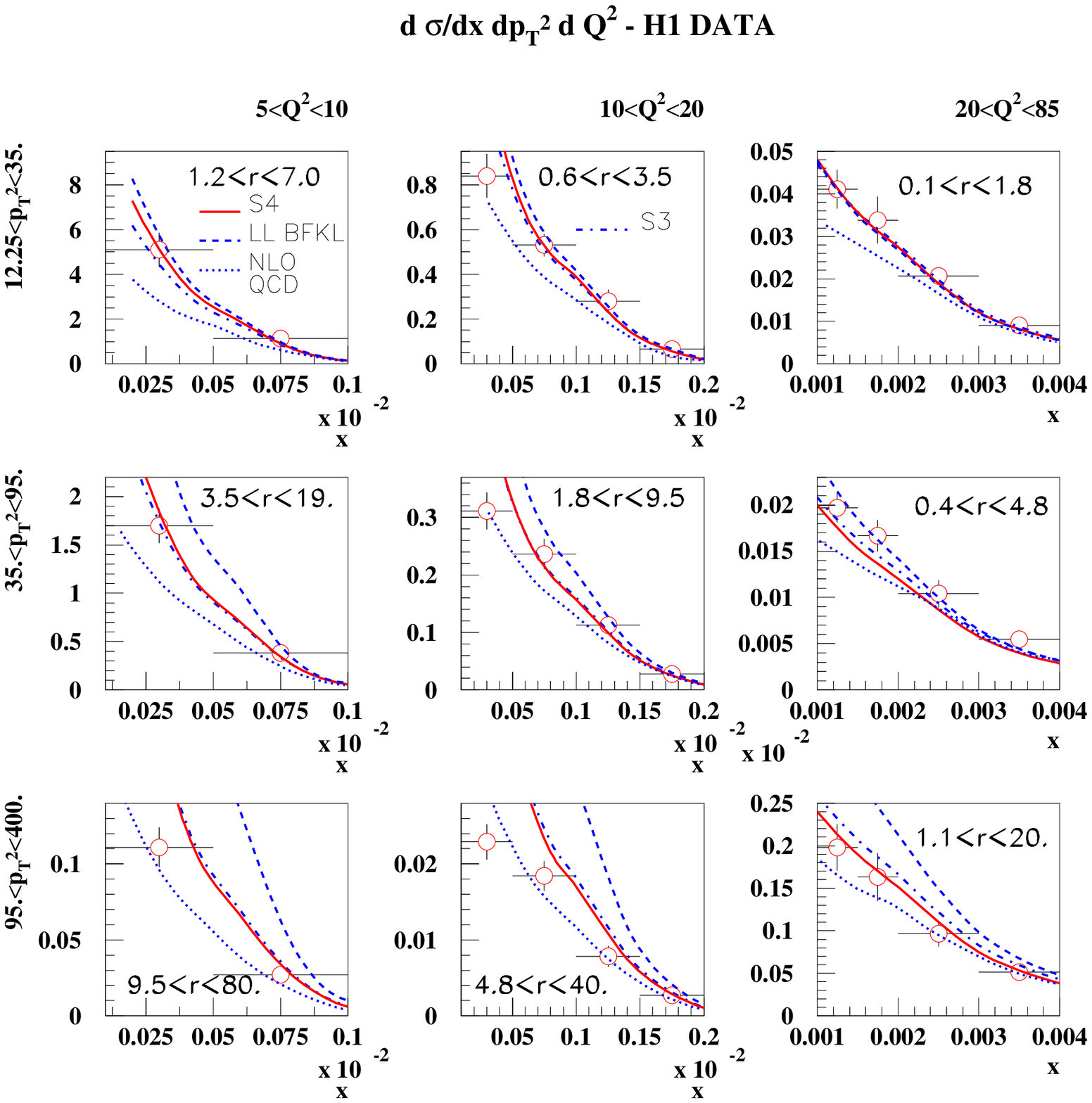,width=12cm}
\caption{The forward-jet cross sections $d\sigma/dx$ (upper plot) and $d\sigma/dxdQ^2dk_T^2$ (lower plot, in nb/GeV$^4$) measured by the H1 collaboration. Comparison with the two NLL-BFKL parametrizations S4 and S3 using the $k_TQ$ scale, and with the LL-BFKL and NLOQCD predictions. In the case of $d\sigma/dx,$ we see a good agreement between the data and the BFKL fits (the NLL-BFKL fits and the LL-BFKL fit are barely distinguishable on the figure) while the NLOQCD predictions do not describe the data. For $d\sigma/dxdQ^2dk_T^2,$ the best description of the data over the whole kinematic range is obtained in the NLL-BFKL approach.}
\end{center}
\end{figure}
\clearpage

\subsection{The triple differential cross section $d\sigma/dxdk_T^2dQ^2$}

The triple differential cross section $d\sigma/dxdk_T^2dQ^2$ is an interesting 
observable as it has been measured with three different $k_T^2$ and $Q^2$ cuts, 
yielding nine different regions for the ratio $r\!=\!k_T^2/Q^2.$ It was noticed in 
\cite{fjetsll2} that the LL-BFKL formalism leads to a good description of the 
data when $r$ is close to 1 and deviates from the data when $r$ is away from 1, as effects
due to the ordering between $Q$ and $k_T$ start to set in. NLOQCD predictions show the reverse trend.

The H1 data for $d\sigma/dxdk_T^2dQ^2$ are shown in Fig.3 (lower plot), and they are compared
with the S4 and S3 predictions, the LL-BFKL results (taken directly from 
\cite{fjetsll2}) and NLOQCD calculations. It is quite remarkable that the 
NLL-BFKL calculation, which includes some ordering between $Q$ and $k_T,$ leads 
to a good description of the H1 data over the full range. As was the case for 
$d\sigma/dx,$ the difference between the LL and NLL results is small when 
$r\!\sim\!1.$ By contrast when $r$ differs from 1, the difference is significant, 
and the observable $d\sigma/dxdk_T^2dQ^2$ is quite sensitive to NLL-BFKL effects.
As a result, the best overall description of the data is obtained with the 
NLL-BFKL formalism. 

\section{Renormalization scale dependence of the NLL description}

\begin{figure}[t]
\begin{center}
\epsfig{file=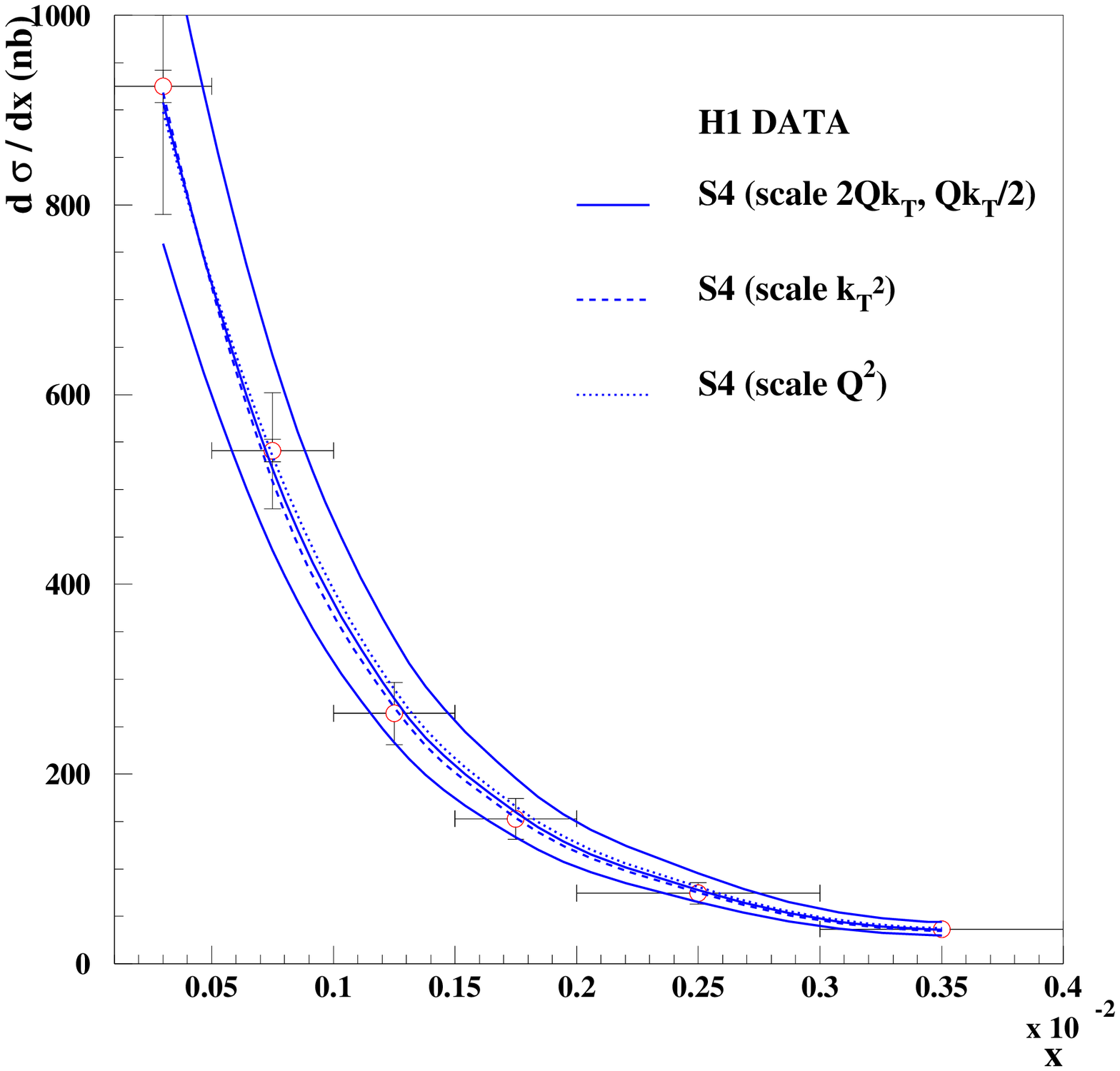,width=6cm}
\epsfig{file=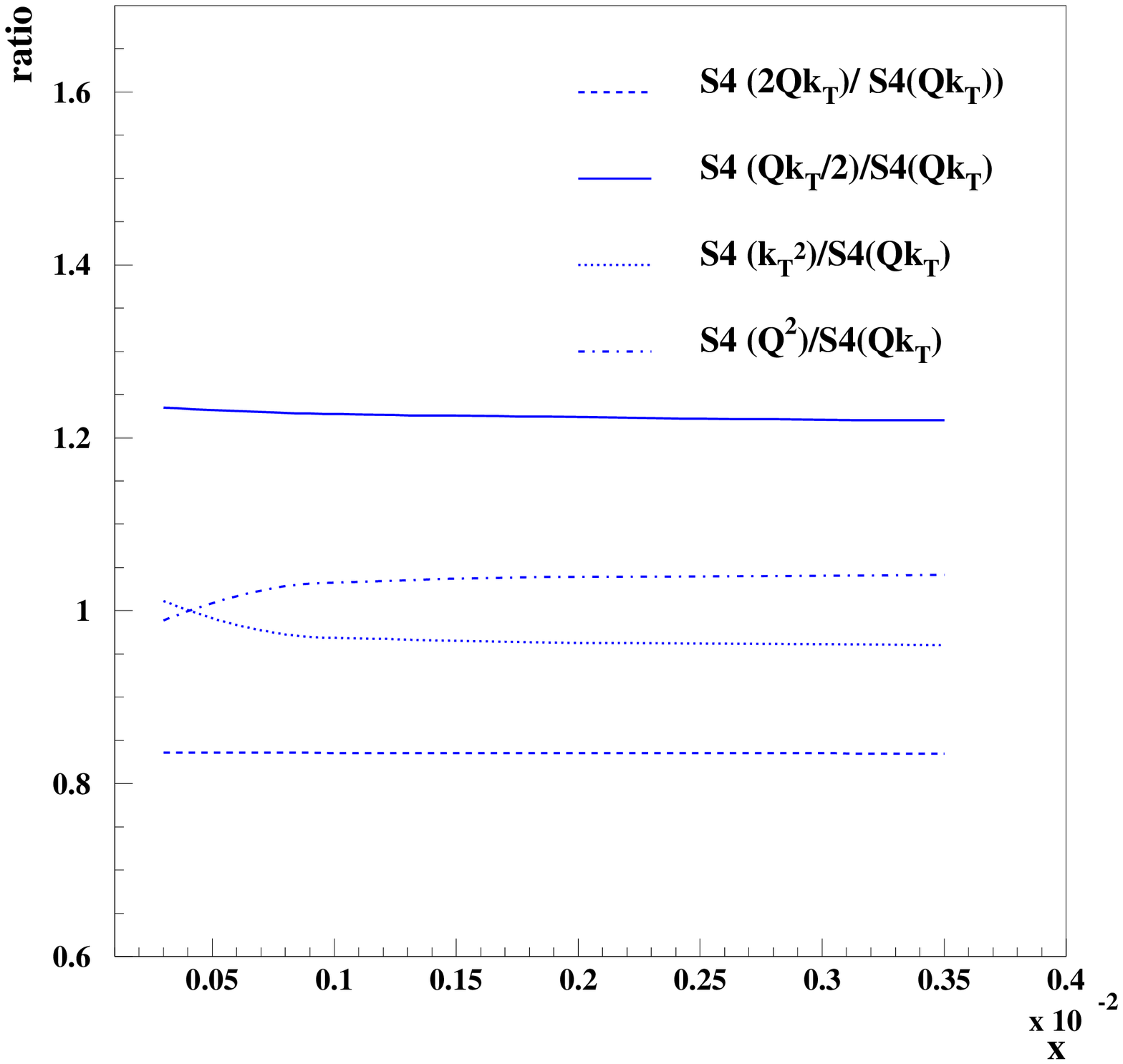,width=6cm}
\epsfig{file=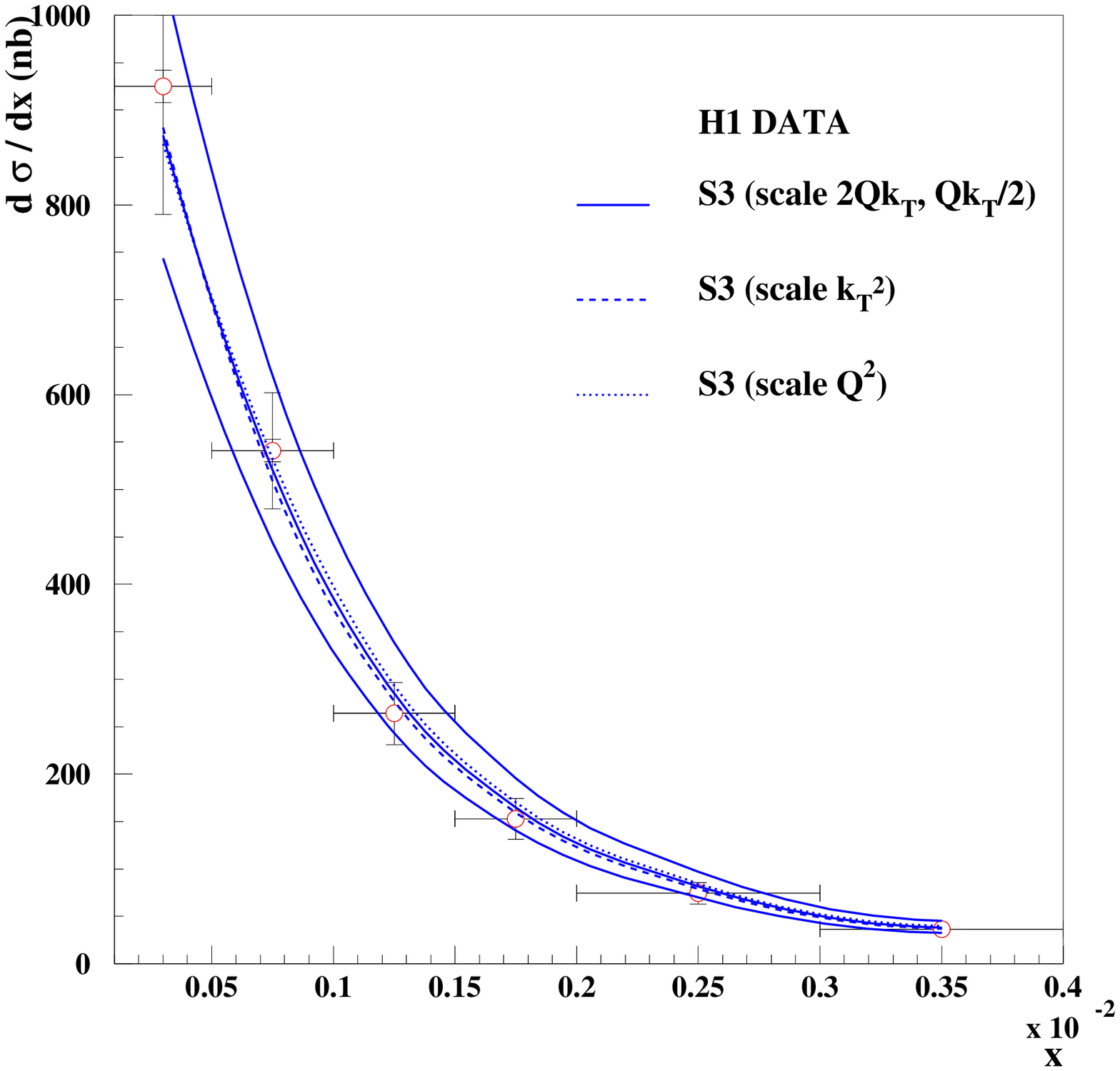,width=6cm}
\epsfig{file=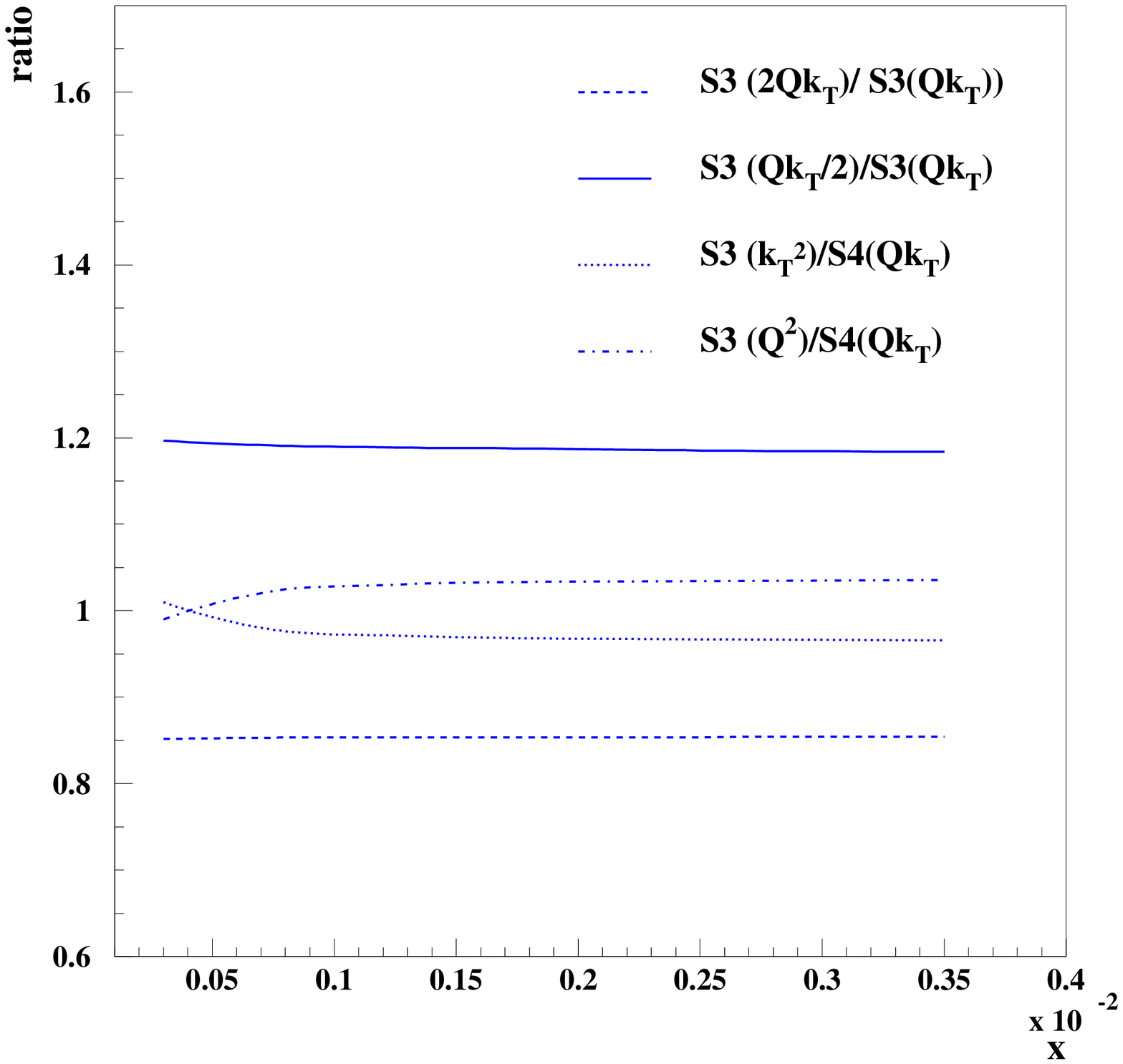,width=6cm}
\caption{
Renormalization scale dependence for the $d\sigma/dx$ cross section. Upper plots: 
S3 scheme, lower plots: S4 scheme. The left plots show the results of the $2k_TQ$ 
and $k_TQ/2$ scales. The right plots show that in the considered range of 
renormalization scale, deviations from the scale $k_T Q$ essentially affect the 
overall normalization. Therefore the quality of the fits (which only adjust the 
normalization) discussed in Section III is not altered by renormalization scale 
uncertainties.}
\end{center}
\end{figure} 

In this section, we study the renormalization-scale dependence of the NLL-BFKL 
description. In the previous section, the choice was $k^2\!=\!k_TQ$ and we now 
test the sensitivity of our results when using $k^2\!=\!\lambda\ k_TQ$ with 
$\lambda\!=\!2,$ $\lambda\!=\!1/2,$ $\lambda\!=\!k_T/Q,$ or $\lambda\!=\!(Q/k_T).$
We use formula \eqref{nll} with the appropriate substitution \cite{modrs}
\be
\bar\alpha(k_TQ)\!\rightarrow\!
\bar\alpha(\lambda k_TQ)\!+\!b\ \bar\alpha^2(k_TQ)\log(\lambda)
\ee
and with the effective kernel modified accordingly following formula 
\eqref{eff}. We also modify the energy scale $k_T Q\!\rightarrow\!\lambda\ k_TQ.$

\begin{figure}[t]
\begin{center}
\epsfig{file=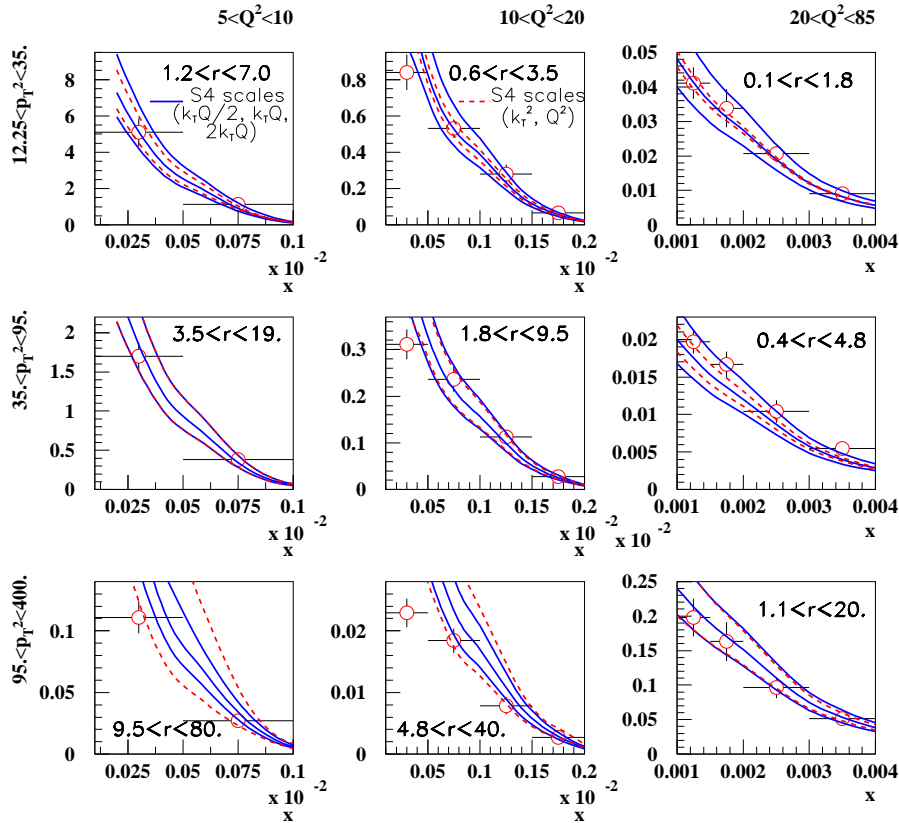,width=12cm}
\caption{Renormalization-scale dependence of the triple differential cross 
section $d\sigma/dxdk_T^2dQ^2$ (in nb/GeV$^4$) for the S4 scheme.}
\end{center}
\end{figure}

We first study the case of $d\sigma/dx$ and the results for the S3 and S4 schemes
are shown in Fig.4. We also display the results in terms of ratios with the
prediction of the $k_T Q$ scale chosen as the reference. We notice that the change of scale essentially affects the overall normalization and thus does not alter the quality of the fit, after readjusting the normalization. This is confirmed by the results of Table II (left table), which presents new fits performed to the $d\sigma/dx$ data for the two scales
$2\ k_TQ$ and $k_TQ/2:$ for each scheme, the $\chi^2$ values are almost insensitive to 
the scale. Note that, for larger values of $\lambda$ such as $\lambda\!=\!4,$ the 
quality of the fit deteriorates due to the large values reached by $\bar\alpha$ in 
the non-perturbative range. Finally, due to the cut $0.5<k_T^2/Q^2<5$ used for the 
measurement \cite{h1new}, using the scales $k_T^2$ or $Q^2$ (instead of $k_TQ$)
yields much less modifications and changes in $\chi^2.$

The next step is to study the effect of the renormalization-scale dependence on 
the triple differential cross section. Applying the normalizations of Table II
while computing the corresponding predictions gives, for either scale, results quite
similar to those of Fig.3. In Fig.5, we rather show the predictions of the
S4 scheme with the common normalization of Table I, and the scale dependence is generally
found to be small. The biggest effects are uncertainties of about the same magnitude as 
the experimental errors, and they are obtained for large values of $r\!=\!k_T^2/Q^2.$
The conclusion is identical in the case of the S3 scheme.

\begin{table}[b]
\begin{center}
\begin{tabular}{|c||c|c|c|} \hline
Scheme & Scale & $\chi^2/dof$ & $N$ \\ 
\hline\hline
    & $k_TQ$  & 10.0/5 & 1.374 $\pm$ 0.016 \\
S4  & $2k_TQ$    & 9.8/5 & 1.644 $\pm$ 0.019  \\      
    & $k_TQ/2$    & 8.8/5 & 1.118 $\pm$ 0.013       \\  
\hline    
\end{tabular}
\hspace{2cm}
\begin{tabular}{|c||c|c|c|c|} \hline
Scheme & Impact factor & $\chi^2/dof$ & $N$ \\
\hline\hline
    &  $\phi^{\g}(\g)$   & 10.0/5 & 1.374 $\pm$ 0.016  \\
S4  &  $\phi^{\g}(1/2)$   & 82.6/5      & 0.655 $\pm$ 0.008  \\
    &  $\phi_{egk}^{\g}(\g,\omega)$ & 23.0/5  &  2.694 $\pm$ 0.038  \\       
\hline
\end{tabular}
\end{center}
\caption{Impact on the fits to the H1 $d\sigma/dx$ data for the S4 scheme when using different renormalization scales (left table) or modified impact factors (right table).}
\end{table}

\section{Impact factor dependence of the NLL description}

\begin{figure}[t]
\begin{center}
\epsfig{file=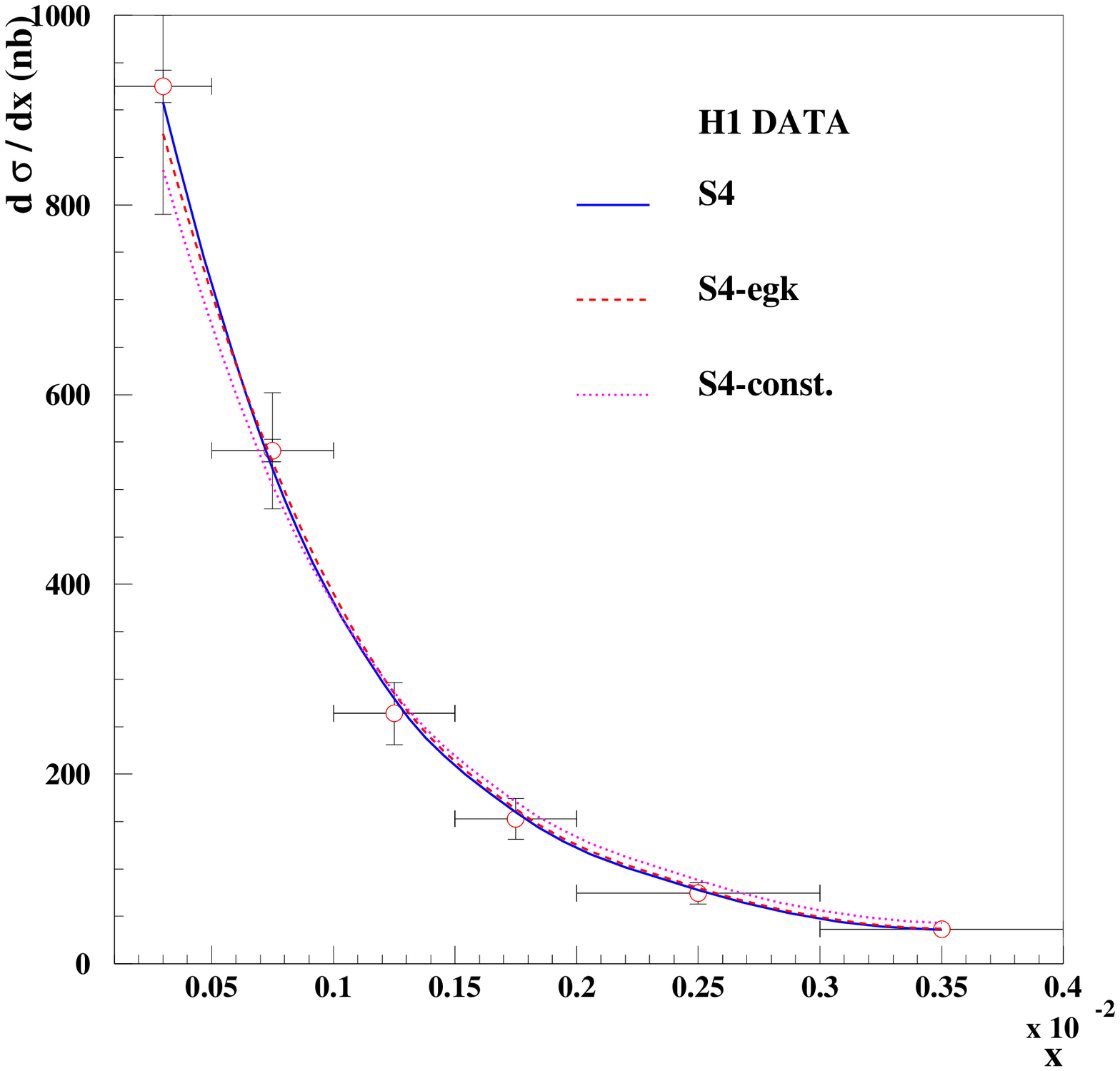,width=6cm}
\epsfig{file=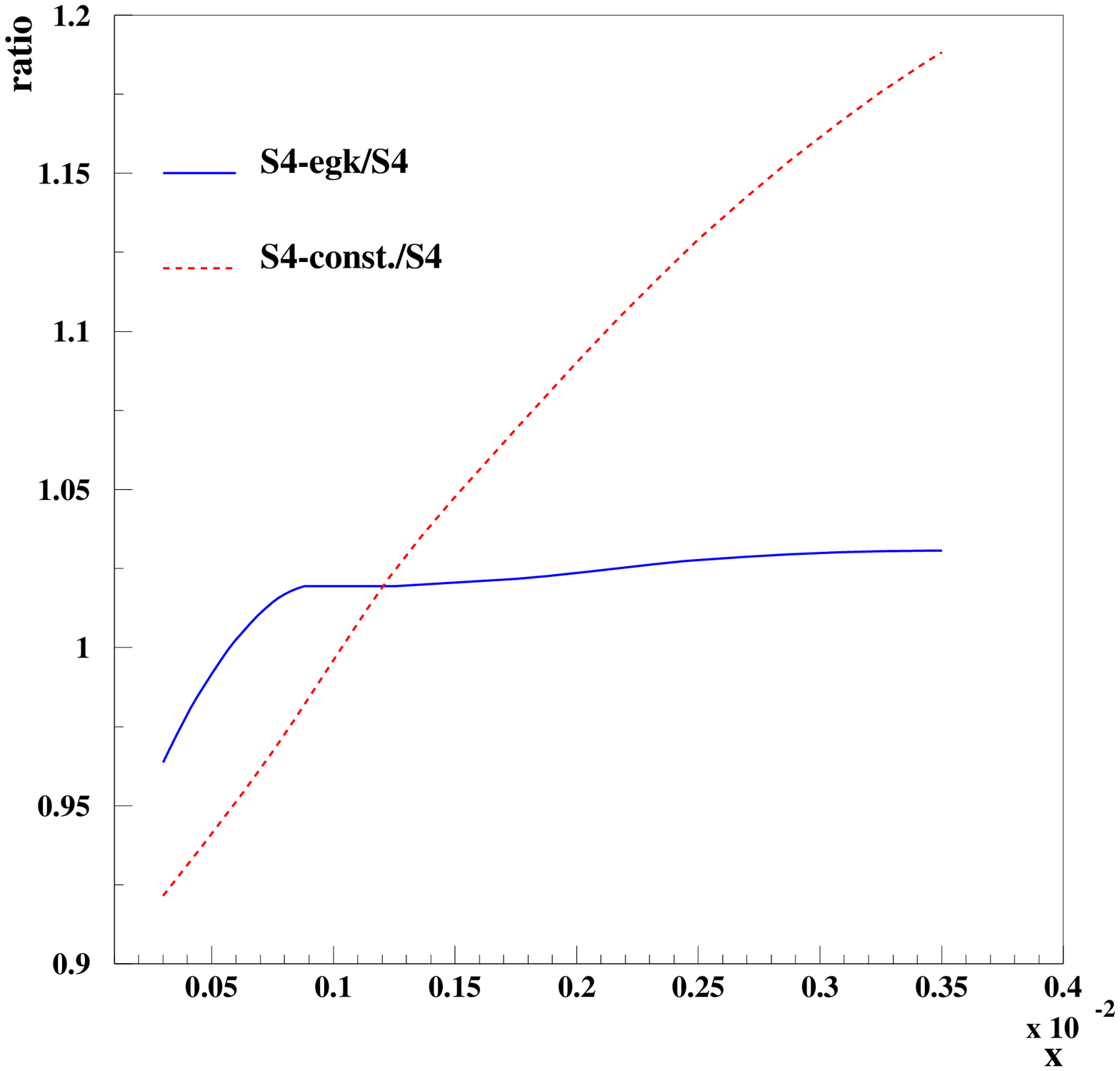,width=6cm}
\caption{Impact factor dependence for the $d\sigma/dx$ cross section in the S4 scheme.
The left plot shows the results of the fits. The right plot shows that in the exact-gluon-kinematics case, the change of impact factors essentially affects the overall normalization, except at very small $x,$ and thus modifies only slightly the quality of the fit.
By contrast in the $\phi^{\g}(1/2)$ case, the $\chi^2$ values are 
sensitive to modifications of the impact factor.}
\end{center}
\end{figure} 

In this section, we study the dependence of our results on typical 
next-leading modifications of the impact factors. Indeed, since the next-leading 
impact factors are unknown, it is useful to see if our results are stable for 
different possible modifications. From \eqref{fj} and \eqref{nll} one notes that 
the impact factors are involved in the computation of the fully differential 
cross section via the following factor in the integrand:
\be
(1-y+y^2/2)\phi^\g_T(\g)+(1-y)\phi^\g_L(\g)\ .\label{impfac}
\ee
In the previous section, the impact factors $\phi^\g_T$ and $\phi^\g_L$ were 
computed as functions of $\g,$ and we treated unknown NLO corrections as a constant 
parameter. We now test the sensitivity of our results when using other prescriptions.

Our first choice is to compute the LO impact factors at $\g\!=\!1/2$ and factor them out
of the $\g$ integration as well. Another prescription that we shall study is the following: implement the higher-order corrections into the impact factor that are due to the exact gluon kinematics in the $\g^*\!\to\!q\bar q g$ transition. These have been calculated in
\cite{bnp} and can be taken into account in our approach: this is done using the
$\omega-$dependent Mellin-transformed impact factors $\phi_{egk}^\g(\g,\omega),$ where we recall that $\omega$ is the Mellin variable conjugate to $W^2/Qk_T.$
Denoting $\delta\!=\!\omega\!-\!2\g\!+\!1,$ one writes:
\bea
\lr{\begin{array}{cc}\phi^\g_{T}(\g,\omega)\\ \phi^\g_{L}(\g,\omega)\end{array}}
=\pi\alpha_{em}N_c^2\sum_q e_q^2\f{1}{4\g^2}
\f{\Gamma(\g\!+\!\delta)\Gamma(\g)}{\Gamma(\omega)(4\!-\!\delta^2)(\delta^2\!-\!1)}
\left[\lr{\begin{array}{cc}3(\omega\!+\!1)^2\!+\!\delta^2\!+\!8
\\24(\g\!+\!\delta)\g\end{array}}\right.\nn\\\left.
-\f{\psi(\g\!+\!\delta)\!-\!\psi(\g)}{2\omega\delta}
\lr{\begin{array}{cc}\omega^2[3(\omega\!+\!1)^2\!+\!9]\!+\!(\delta^2\!-\!1)
(\delta^2\!-\!2\omega\!-\!9)\\8(\g\!+\!\delta)\g(3\omega^2\!-\!\delta^2\!+\!1)\end{array}}
\right]\ .\label{egk}
\eea
In practice, following \eqref{omegaint}, these impact factors are evaluated at 
$\omega\!=\!\bar\alpha(k_T Q)\chi_{eff}[\g,\bar\alpha(k_T Q)]$ prior to the $\gamma$ integration.
This prescription is motivated by the fact that, in the proton structure function analysis, these higher-order corrections allow for an improved DGLAP analysis \cite{white}: indeed, they match the fixed-order results (at NLO and NNLO) for the splitting and coefficient functions of the DGLAP approach.

We first study the case of $d\sigma/dx.$ Table II (right table) presents new fits 
performed to the $d\sigma/dx$ data for the different prescriptions. The fit results are shown in Fig.6 (left plot) and we also display them in terms of ratios with the prediction of the
$\phi^{\g}(\g)$ case chosen as the reference (right plot). In the $\phi_{egk}^\g(\g,\omega)$ case, we notice that the change of impact factors essentially affects the overall normalization, except at very small values of $x$ where the gluon kinematics become more restrictive. As a result the quality of the fit, which readjusts the normalization, is only slightly modified. By contrast in the $\phi^{\g}(1/2)$ case, the shape is clearly different, which yields a bad
$\chi^2.$

After applying the normalizations of Table II (right table), we now compare the
effect of the different impact factors on the triple differential cross section.
The results are given in Fig.7. The effect is found to be small in the 
exact gluon-kinematics case, and even slightly improves the description. By contrast, 
differences are important in the $\phi^{\g}(1/2)$ case, especially at large values of
$r\!=\!k_T^2/Q^2.$

\begin{figure}[t]
\begin{center}
\epsfig{file=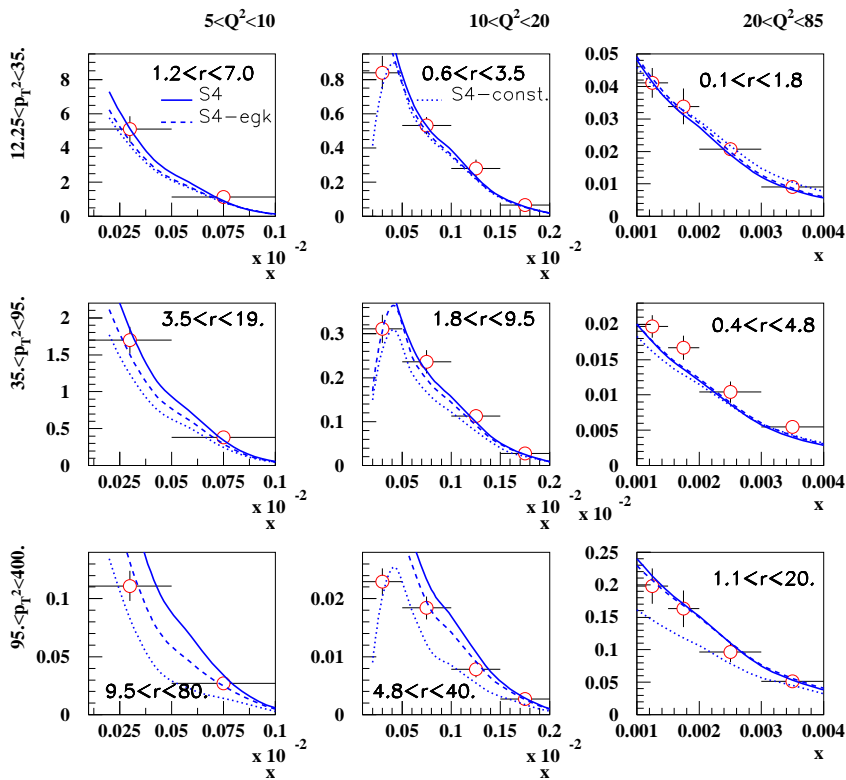,width=11.3cm}
\caption{Impact factor dependence of the triple differential cross section $d\sigma/dxdk_T^2dQ^2$ (in nb/GeV$^4$) in the S4 scheme. The relative normalizations coming from the $d\sigma/dx$ fits have been applied.}
\end{center}
\end{figure}

\section{PDF and scale dependence of the NLOQCD predictions}

To obtain the NLOQCD predictions of the forward-jet data, the photon-parton hard 
cross section is computed at next-to-leading order in $\alpha_s,$ and the leading and next-leading logarithms $\alpha_s^n\log^n{Q^2}$ 
and $\alpha_s^n\log^{n-1}Q^2$ are resummed using the DGLAP equations
\cite{dglap} which govern the evolution of the parton distribution functions 
(PDFs) of the proton.

The prediction for the forward-jet cross section at next-to-leading order is 
calculated using NLOJET++ \cite{nlojetDIS}. The renormalization scale $\mu_r^2$ 
and the factorization scale $\mu_f^2$ were chosen as 
$\mu_r^2=\mu_f^2=Qk_T^{max},$ where $k_T^{max}$ denotes the maximal transverse 
momentum of forward jets in the event. To obtain the uncertainty associated with 
the choice of the scale, we varied the scales in the conventional range 
$Qk_T^{max}/4\!<\!\mu_r^2\!=\!\mu_f^2\!<\!4Qk_T^{max}$. Another scale choice 
$\mu_r^2\!=\!\mu_f^2\!=\!k_T^{max2}$ was also tested; however, it yielded a 
result located within the mentioned scale uncertainty bounds and thus shall not 
be considered further. We point out that the scale uncertainty is rather large in 
the low$-x$ regime. This is due to the large NLO correction suggesting that 
higher-order corrections might be significant. 

An additional contribution to the total uncertainty of the calculation is due to 
the choice of the PDFs of the proton. Throughout the calculation, we use the 
CTEQ6.1M PDF parametrization which provides not only the central PDF $S_0$ 
corresponding to the best fit to data, but also 40 additional distribution 
functions $S_i^+$, $S_i^-$, $i=1\dots20$ devoted to uncertainty studies
\cite{cteq}. The total PDF uncertainty $\Delta X$ of the observable $X$ is then 
computed as
\be
(\Delta X)^2=\sum\limits_{i=1}^{20}\left[\frac{X(S_i^+)-X(S_i^-)}{2}\right]^2\ .
\ee
We noticed that the main contribution to the PDF uncertainty comes from the gluon 
PDF.

The DGLAP calculation of the $d\sigma/dx$ distribution measured by H1 is 
presented in Fig.8, upper plot. This approach clearly fails to describe the data 
for the low values of $x.$ In comparison with Fig.3, we see that 
$d\sigma/dx$ is more sensitive to the BFKL dynamics than the triple differential 
distribution as the deviation from the data is obvious.

\clearpage
\begin{figure}[t]
\begin{center}
\epsfig{file=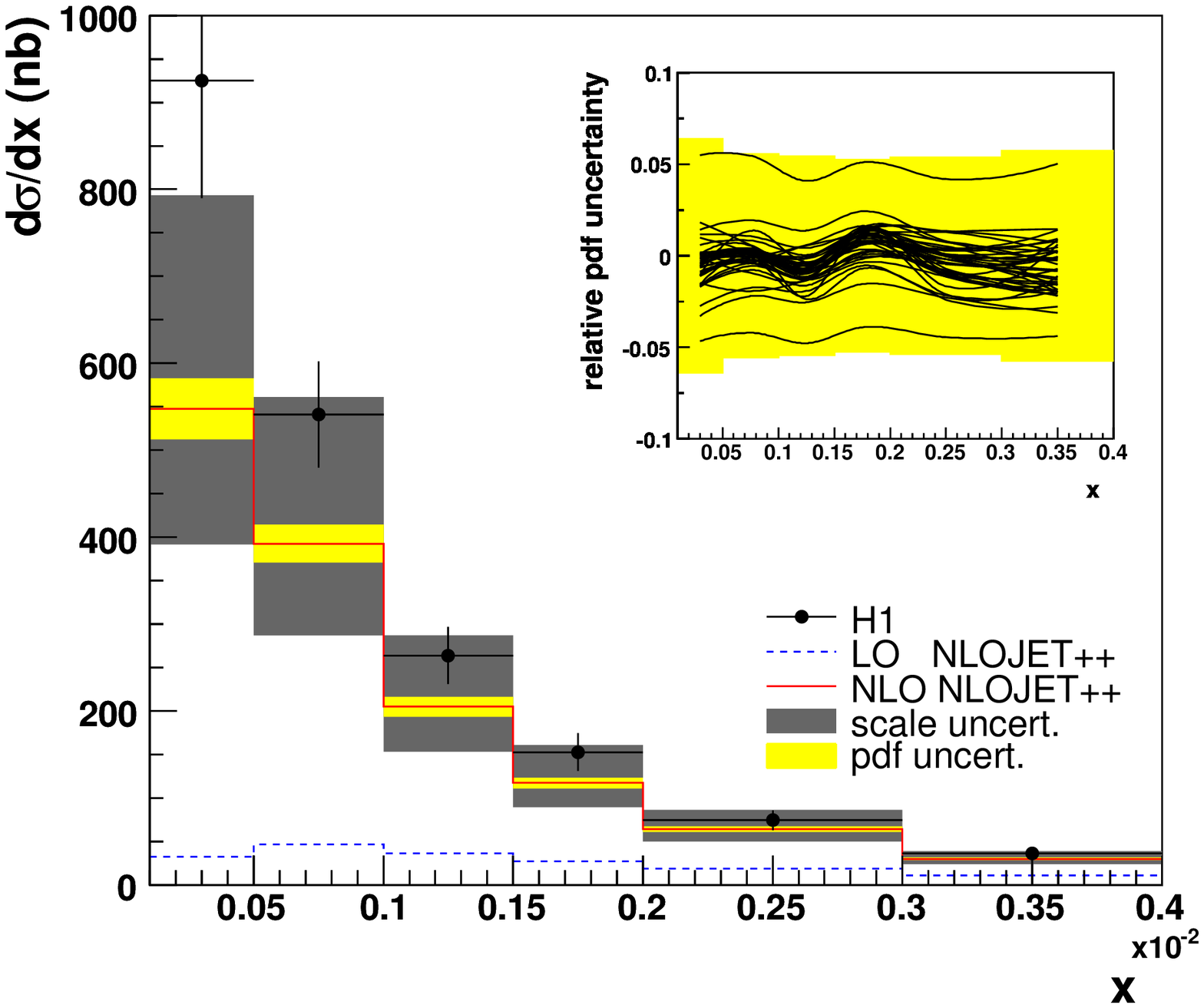,width=8.5cm}
\epsfig{file=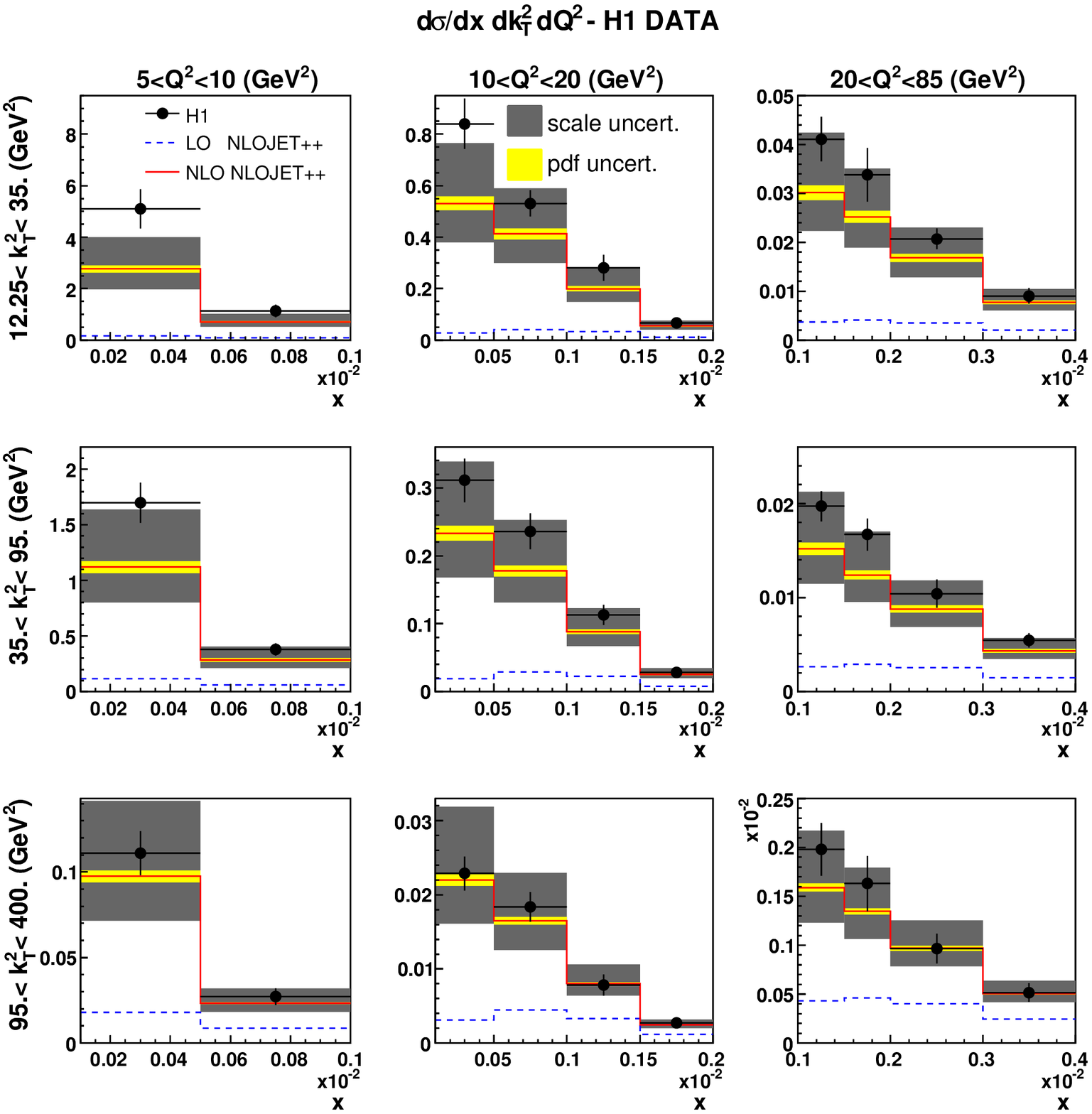,width=13cm}
\caption{Upper plot: scale and PDF uncertainties of the $d\sigma/dx$ 
distribution; the particular contributions of 40 different PDFs from CTEQ6.1M to 
the total PDF uncertainty (yellow area) is depicted in the inset. Lower plot: 
scale and PDF uncertainties for the triple differential distribution 
$d\sigma/dxdQ^2dk_T^2$ (in nb/GeV$^4$). In both cases, the scales varying in the range
$0.25\ Qk_T^{max}<\mu_r^2=\mu_f^2<4\ Qk_T^{max}$ are displayed (shaded gray area).}
\end{center}
\end{figure} 
\clearpage

The PDF uncertainty study is shown in the inset. The solid lines represent the ratio 
of the cross section calculated with the various PDFs $S_i^+$, $S_i^-$, normalized to 
the cross section calculated with the central PDF $S_0$. The gluon PDF (i=15) has 
the greatest impact on the uncertainty and the other PDFs can be neglected. The 
uncertainty study of $d\sigma/dxdk_T^2dQ^2$ is shown in Fig.8, lower plot. The PDF 
uncertainty was obtained taking into account the gluon density (which practically 
dictates the overall PDF uncertainty) only.

The main conclusion that can be drawn from Fig.8 is that at low values of
$x,$ the NLOQCD results suffer from large uncertainties, and they indicate that NNLO 
calculations are needed to obtain genuine predictions in this framework.

\section{Conclusions}

We performed a phenomenological analysis of the H1 forward-jet data, looking for 
effects of next-to-leading logarithms in the BFKL approach. Let us briefly 
summarize our results.

\begin{itemize}

\item For the cross section $d\sigma/dx,$ measured in the kinematical regime
$Q^2/k_T^2\!\sim\!1,$ we obtain a good description of the H1 data by NLL-BFKL 
predictions (see Table I and Fig.3, upper plot). In addition, the difference between the 
LL-BFKL and NLL-BFKL descriptions is very small once the overall normalization is fit. This confirms the validity of the BFKL description of
\cite{fjetsll,fjetsll2} previously obtained with the LL formula and a rather small 
effective coupling. 

\item In the case of the triple differential cross section
$d\sigma/dxdk_T^2dQ^2,$ the same conclusions holds when
$r\!=\!k_T^2/Q^2\!\sim\!1.$ In addition when $r$ differs from 1, the NLL-BFKL 
description is quite different from the LL-BFKL one, as it is closer to the NLOQCD 
calculation. As a result, the best overall description of the data for
$d\sigma/dxdk_T^2dQ^2$ is obtained with the NLL-BFKL formalism (see Fig.3, lower plot).

\item The renormalization-scale dependence of our results has been thoroughly 
studied and we showed the stability of the NLL-BFKL approach when using the scales 
$k_TQ,$ $2k_TQ$ and $k_TQ/2.$ For $d\sigma/dx,$ the change of 
scale essentially affects the overall normalization and does not alter the quality 
of the fits (see Fig.4 and Table II, left table). For $d\sigma/dxdk_T^2dQ^2,$ the biggest 
effect yields an uncertainty of about the same magnitude as the experimental errors 
(see Fig.5).

\item We want to stress that the HERA data allow for a detailed study of the 
NLL-BFKL approach and of the QCD dynamics of forward jets. In particular, it has 
the potential to address the question of the remaining ambiguity corresponding to 
the dependence on the specific regularization scheme of the NLL kernel. For 
instance, the predictions of the S3 scheme do not compare with the data as 
well as the predictions of the S4 scheme, as indicated by the $\chi^2$ values given in
Table I. Therefore, it would be very interesting to compare the data with other
regularization procedures \cite{singnll,modrs} than those used here. However,
these other solutions proposed to remove the spurious singularities of the NLL
kernel are not in such a suitable form for phenomenology, hence this issue will
be addressed in a separate work.

\item Our analysis is to be completed with the next-leading photon impact factors, 
when available. However we tested the stability of our approach 
when implementing typical next-leading modifications of leading-order impact 
factors. The results show some sensitivity (see Fig.6-7 and Table II, right table)
with the $\phi^{\g}(1/2)$ prescription. It is however interesting that using the impact
factors \eqref{egk} with exact gluon kinematics (which, in the structure function case,
allow for an improved DGLAP analysis) still gives a good fit of the $d\sigma/dx$ data, and also
a better description of the $d\sigma/dxdk_T^2dQ^2$ cross section. This indicates that,
when the next-leading impact factors will be known, our predictions could be stable.

\item Finally, we computed the NLOQCD predictions using the NLOJET++ generator and 
CTEQ6.1M \cite{cteq} parton densities. We tested their relevance by comparing the 
use of different parton densities and renormalization and factorization scales (see 
Fig.8). The NLOQCD predictions do not describe the data at small values of $x,$ 
and they suffer from large uncertainties, showing the need for higher-order 
corrections in this framework.

\end{itemize}

Forward-jet production is the first observable for which the NLL-BFKL description 
works, while the standard NLOQCD does not work. We need complete knowledge of 
the next-to-leading impact factors before drawing final conclusions, but our analysis 
strongly suggests that the data show the BFKL enhancement at small values of $x.$ 
This is of great interest in view of the LHC, where similar QCD dynamics will be 
tested with Mueller-Navelet jets \cite{mnjets,mnj}.

\begin{acknowledgments}

This research was supported in part by RIKEN, Brookhaven National Laboratory and the U.S. Department of Energy [DE-AC02-98CH10886].

\end{acknowledgments}

\section*{APPENDIX I: The S3 and S4 regularization schemes}

In this appendix, we recall the regularization procedure of \cite{salam} to obtain
$\chi_{NLL}\lr{\g,\omega}$ in the S3 and S4 schemes. The starting point is the 
scale invariant (and $\g\!\leftrightarrow\!1-\g$ symmetric) part of the NLL-BFKL 
kernel
\bea
\chi_1(\g)=\f32\zeta(3)+\lr{\f{1+5b}3-\f{\zeta(2)}2}\chi_{LL}(\g)-\f{b}2\chi_{LL}^2
(\g)
+\f14\left[\psi''(\g)+\psi''(1-\g)\right]\nn\\
-\phi(\g)-\f{\pi^2\cos(\pi\g)}{4\sin^2(\pi\g)(1-2\g)}
\left[3+\lr{1+\f{N_f}{N_c^3}}\f{2+3\g(1-\g)}{(3-2\g)(1+2\g)}\right]\eea
with $b$ given in \eqref{runc}, $\chi_{LL}$ given in \eqref{nlltoll}, and 
\be
\phi(\g)=\f12\sum_{k=0}^\infty\f{k+1/2}{(k+\g)(k+1-\g)}
\left[\psi'\lr{\f{k+2}2}-\psi'\lr{\f{k+1}2}\right]\ .
\label{klcor}\ee
The pole structure of $\chi_1(\g)$ at $\g=0$ (and by symmetry at $\g=1$) is:
\be
\chi_1(\g)=-\f1{2\g^3}+\f{d_2}{\g^2}+\f{d_1}{\g}+{\cal O}(1)
\ee
with
\be
d_1=-\f{N_f}{18N_c}\lr{5+\f{13}{2N_c^2}}\ ,\hspace{0.5cm}
d_2=-\f{11}8+\f{N_f}{12N_c}\lr{1-\f{2}{N_c^2}}\ .
\ee
\begin{itemize}
\item The S3 scheme kernel $\chi_{S3}\lr{\g,\omega}$ is given by 
\bea
\chi_{S3}(\g,\omega)=[1-\bar\alpha A]
\left[2\psi(1)-\psi\lr{\g+\f{2\bar\alpha B+\omega}2}
-\psi\lr{1-\g+\f{2\bar\alpha B+\omega}2}\right]
\nn\\+\bar\alpha\left\{\chi_1(\g)+A\chi_{LL}(\g)
+\lr{B+\f{\chi_{LL}(\g)}2}\left[\psi'(\g)+\psi'(1-\g)\right]\right\}
\eea
with $A$ and $B$ chosen to cancel the singularities of $\chi_1(\g)$ at $\g=0:$
$A=-d_1-\pi^2/6$ and $B=-d_2.$
\item The S4 scheme kernel $\chi_{S4}\lr{\g,\omega}$ is defined with the help of 
the function
$f(\g)=1/\g+1/(1-\g):$
\be
\chi_{S4}(\g,\omega)=\chi_{LL}(\g)-f(\g)+[1-\bar\alpha A]f(\omega+2\bar\alpha B,\g)
+\bar\alpha\left\{\chi_1(\g)+Af(\g)+\lr{B+\f{\chi_{LL}(\g)}2}
\left[\f1{\g^2}+\f1{(1-\g)^2}\right]\right\}\ .
\ee
In this scheme, $A$ and $B$ are given by $A=-d_1-1/2$ and $B=-d_2.$
\end{itemize}

\section*{APPENDIX II: Comparison between the exact NLL-BFKL $\g$ integration and a 
saddle-point approximation}

It is possible to estimate the complex integration in \eqref{nll} using a 
saddle-point approximation in $\g.$ In the BFKL regime we are working in, $Y$ is 
very large, and the saddle-point equation 
\be
\f{d\chi_{eff}}{d\g}(\g_c,\bar\alpha)=\chi'_{eff}(\g_c,\bar\alpha)=
\f{\log(k^2_T/Q^2)}{\bar\alpha Y}
\ee
becomes $\chi'_{eff}(\g_c,\bar\alpha)\!=\!0.$ Hence one finds for the theoretical 
forward-jet cross section
\be
\f{d\sigma^{\g*p\!\rightarrow\!JX}_{T,L}}{dx_Jdk_T^2}\simeq
\f{\alpha_s(k_T^2)\alpha_s(Q^2)}{k_T^2Q^2}f_{eff}(x_J,k_T^2)\
\lr{\f{Q^2}{k_T^2}}^{\g_c}\f{\phi^\g_{T,L}(\g_c)}
{\sqrt{2\pi\bar\alpha\chi_{eff}''(\g_c,\bar\alpha)\ Y}}\ 
\exp\lr{\D\bar\alpha\chi_{eff}(\g_c,\bar\alpha)Y
-\f{\log^2(Q^2/k_T^2)}{2\bar\alpha\chi_{eff}''(\g_c,\bar\alpha)\ Y}}\ ,
\label{nllsaddle}\ee 
where $\chi_{eff}''\!=\!d^2\chi_{eff}/d\g^2.$ 

\clearpage

\begin{figure}[t]
\begin{center}
\epsfig{file=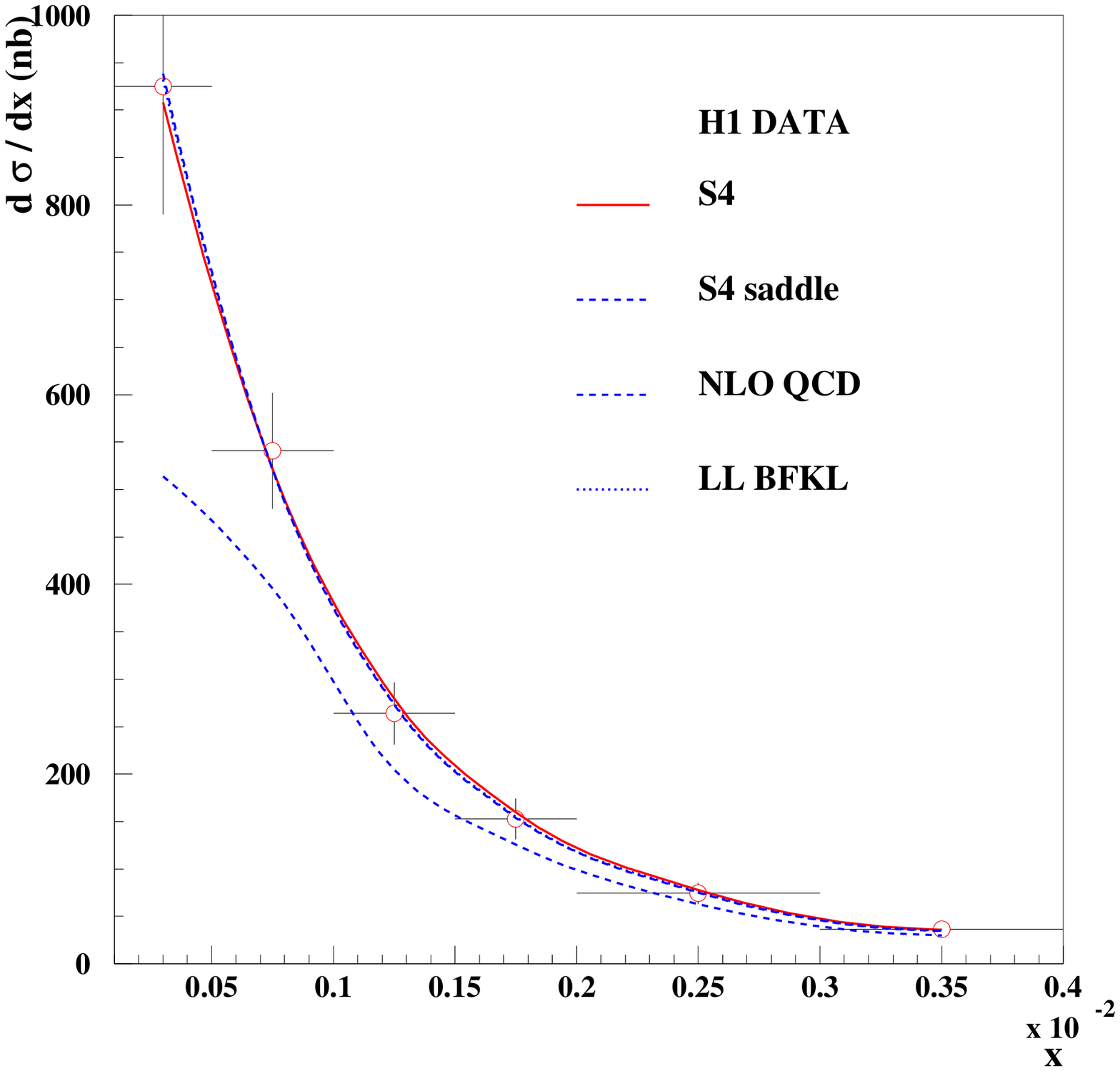,width=9cm}
\epsfig{file=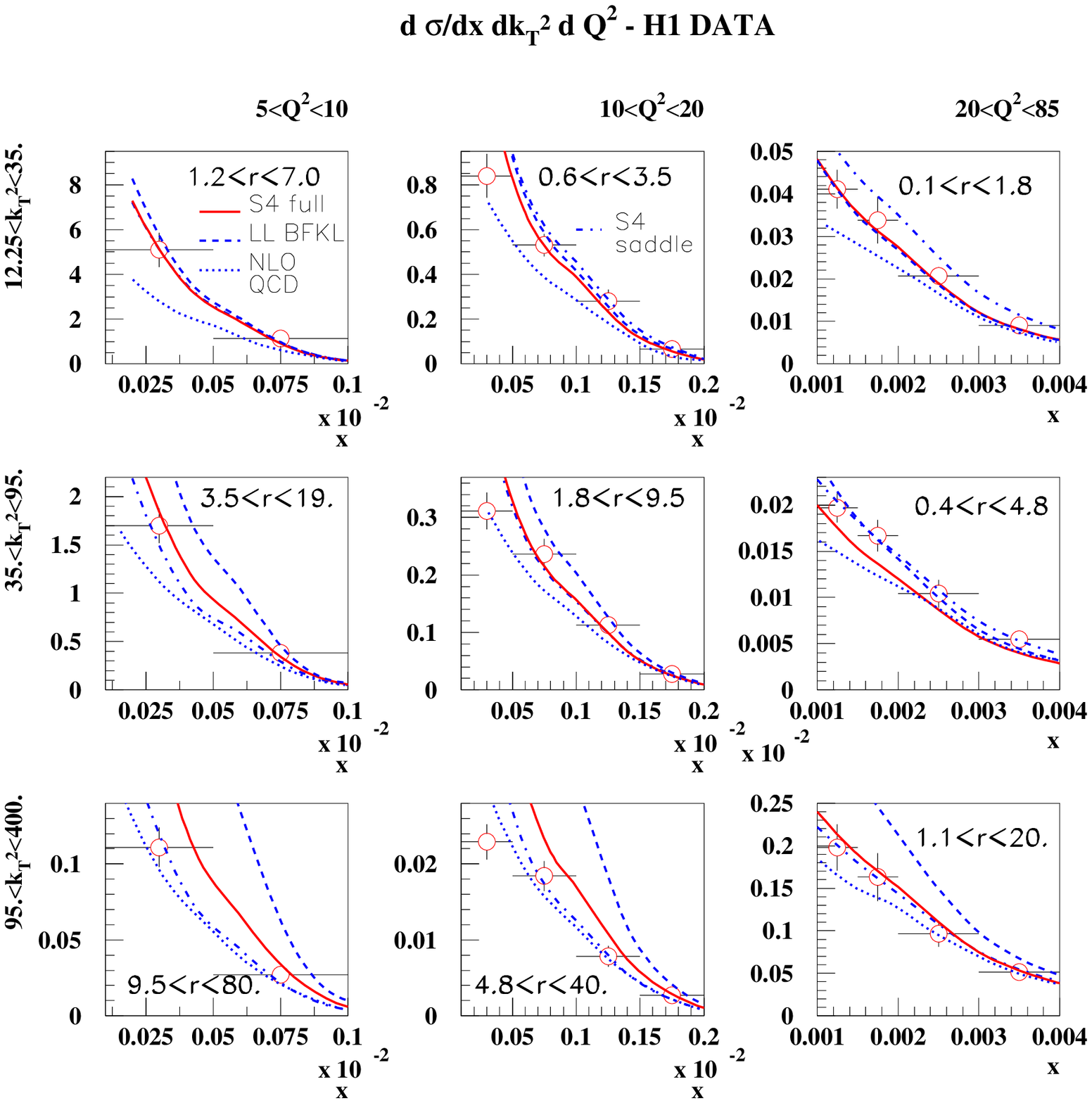,width=12.5cm}
\caption{The forward-jet cross sections $d\sigma/dx$ (upper plot) and $d\sigma/dxdQ^2dk_T^2$ 
(lower plot, in nb/GeV$^4$) measured by the H1 collaboration and compared with two NLL-BFKL predictions of the S4 scheme, obtained with the exact $\g$ integration \eqref{nll} or the saddle-point approximation 
\eqref{nllsaddle}. The LL-BFKL and NLOQCD results are also recalled.}
\end{center}
\end{figure} 
\clearpage

For each set of scales ($Q^2,k_T^2$), it is possible to extract the 
values of $\g_c,$ $\bar\alpha\chi_{eff}(\g_c,\bar\alpha)$ and 
$\bar\alpha\chi''_{eff}(\g_c,\bar\alpha)$ after solving the implicit equation 
\eqref{eff}. 

This approach was considered in \cite{us}, and compared with the results obtained 
with the exact integration. In this appendix we discuss this comparison in more 
detail. At the level of the differential cross sections \eqref{nll} and 
\eqref{nllsaddle} there are some differences between the exact NLL-BFKL integration 
and the saddle-point approximation. But when considering the integrated, 
experimentally measured cross sections $d\sigma/dx$ and 
$d\sigma/dxdk_T^2dQ^2,$ the description of the data is similar. This is shown in 
Fig.9.

We recall that, starting from \eqref{fj}, one has to carry out a number of 
integrations over the kinematic variables, which have to be done while properly 
taking into account the kinematic cuts applied for the different measurements 
\cite{fjetsll2}. This procedure seems to erase the differences between the exact 
NLL-BFKL integration and the saddle-point approximation. This is even more so 
for $d\sigma/dx.$


\end{document}